\documentclass[%
 reprint,
 amsmath,amssymb,
 aps,
 prb,
]{revtex4-1}

\usepackage{graphicx}
\usepackage{dcolumn}
\usepackage{bm}
\usepackage{color}
\usepackage{times}
\usepackage{amsmath}
 
\begin{document}


\title{Demultiplexing infrasound phonons with tunable magnetic lattices}

\author{Audrey A. Watkins and Osama R. Bilal} 
\affiliation{Department of Mechanical engineering, University of Connecticut, Storrs, USA}

\date{\today}
\begin{abstract}

Controlling infrasound signals is crucial to many processes ranging from predicting atmospheric events and seismic activities to sensing nuclear detonations. These waves can be manipulated through phononic crystals and acoustic metamaterials. However, at such ultra-low frequencies, the size (usually on the order of meters) and the mass (usually on the order of many kilograms) of these materials can hinder its potential applications in the infrasonic domain. Here, we utilize tunable lattices of repelling magnets to guide and sort infrasound waves into different channels based on their frequencies. We construct our lattices by confining meta-atoms (free-floating macroscopic disks with embedded magnets) within a magnetic boundary. By changing the confining boundary, we control the meta-atoms’ spacing and therefore the intensity of their coupling potentials and wave propagation characteristics. As a demonstration of principle, we present the first experimental realization of an infrasound phonon demultiplexer (i.e., guiding ultra-low frequency waves into different channels based on their frequencies). The realized platform can be utilized to manipulate ultra-low frequency waves, within a relatively small volume, while utilizing negligible mass. In addition, the self-assembly nature of the meta-atoms can be key in creating re-programmable materials with exceptional nonlinear properties.


\end{abstract}

\maketitle

Infrasound waves are ubiquitous in  nature, appearing in animal communications,\cite{von2003songlike, barklow2004low, von2013giraffe} natural phenomena such as earthquakes and volcanic eruptions,\cite{nakano2018shallow,fee2013overview, ripepe2018infrasonic,garces2003observations} and man made systems such as machinery and explosions. These waves can cause high volumes of damage to their surroundings, particularly when originating from natural phenomena.\cite{campus2010worldwide} Deciphering infrasound waves can lead to potential detection of various natural phenomena prior to their occurrences, offering a type of early warning system through their detection. Phononic crystals and acoustic metamaterials, defined as artificial arrangements of spatial patterns, can manipulate waves at different frequency ranges from a few Hertz to a few Terahertz.\cite{maldovan2013sound, deymier2013acoustic, hussein2014dynamics,khelif2015phononic} These phononic metamaterials usually consist of unit cells that repeat periodically. Phononic crystals have the ability to affect waves through Bragg scattering,\cite{kushwaha1993acoustic} while metamaterials utilize resonances to affect waves.\cite{liu2000locally} The wave controlling characteristics of phononic crystals and acoustic metamaterials have a variety of potential applications, including vibration and sound insulation,\cite{yang2010acoustic,mei2012dark,ma2015purely} seismic wave protection,\cite{kim2012seismic,brule2014experiments} wave guiding,\cite{Torres_1999,rupp2007design} frequency filtering,\cite{rupp2010switchable} acoustic lenses,\cite{moleron2014acoustic} impact absorption, \cite{chen2017wave,barnhart2019experimental} and sensors.\cite{ke2011sub,amoudache2014simultaneous}

The presence of scattering in phononic crystals and resonances in metamaterials gives rise to specific frequency ranges where waves cannot propagate (i.e., band gaps). In a scattering-based band gap, the unit cell size has to be on the same order of the targeted frequency wavelength. This usually translates into a meter sized structure that affects waves at ultra-low frequencies.\cite{brule2014experiments} Resonance based metamaterials can open band gaps where the targeted frequency wavelength does not correlate with the unit cell size (i.e, subwavelength band gaps). However, opening band gaps at ultra-low frequencies could translate to a resonator with mass on the order of many kilograms\cite{palermo2016engineered, colombi2020mitigation} or coupling stiffness that is prone to fatigue. In addition, most phononic metamaterial realizations have fixed operational frequency. Once a sample is fabricated, its band gap frequency, for example, cannot change.\cite{matar2013tunable,bergamini2014phononic,bilal2017reprogrammable} 


In this paper we utilize tunable lattices of repelling magnets to guide and sort infrasound waves into different channels based on their frequencies. We realize these lattices by confining meta-atoms (free-floating macroscopic disks with embedded magnets) within a magnetic boundary. We control the spacing between the meta-atoms by changing the magnetic potential of the confining boundary. The same meta-atoms with different lattice constants and coupling intensity can have different wave propagation characteristics by design. As a demonstration of principle, we experimentally realize an infrasound phonon demultiplexer (i.e., guiding ultra-low frequency waves into different channels based on their frequencies). By constructing channels with identical building blocks, yet different boundary potentials, we can demultiplex infrasound waves based on their frequencies. There exist multiple theoretical proposals for realizing phonon demultiplexers\cite{rostami2016designing, pennec2004tunable, moradi2019three, motaei2020eight,babaki2020heterostructure,ben2020shaped, gharibi2020phononic, zou2017decoupling, pennec2005channel, hussein2005hierarchical, vasseur2011band} with few experimental demonstrations.\cite{mohammadi2011chip,faiz2020experimental,bilal2020experimental} However, demultiplexing infrasound phonons has not been observed.

\begin{figure}[t]
	\begin{center}
		\includegraphics{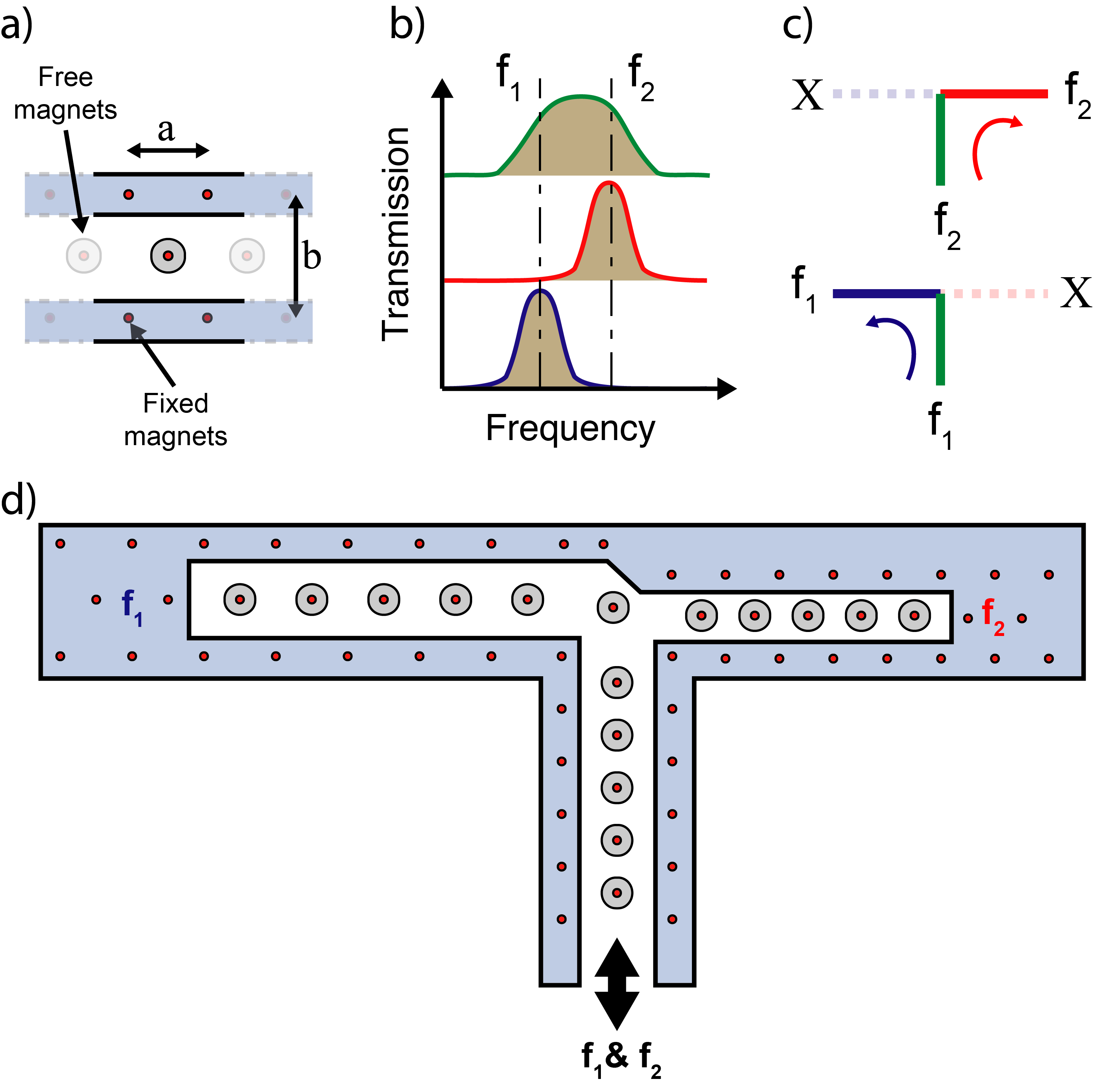}
	\end{center}
	\vspace*{-7mm}
	\caption{ \textbf{Design concept}. a) Schematic of a unit cell depicting the locations of $a$ and $b$ as well as the fixed and free magnets. b) Transmission spectrum of channel 1 (blue), 2 (green), and 3 (red). c) A visual representation of two different frequencies ($f_1$ and $f_2$) passing through the channels. d) A full schematic of the t-shaped demultiplexer with $f_1$ and $f_2$ excited at the vertical channel (CH. 2) and sorted into their respective channels.}
	\label{fig:Schematic}
\end{figure}

To design our infrasound phonon demultiplexer, we start by considering a unit cell composed of a disk with a concentrically embedded magnet, referred to as meta-atoms hereinafter. The meta-atoms are confined between two arrays of fixed, identical, permanent magnets (Fig. \ref{fig:Schematic} a). The fixed magnets create a boundary potential to hold the meta-atoms in place. The fixed magnets are equally spaced by a length $a$. The two arrays of fixed magnets are separated by a distance $b$. Both $a$ and $b$ define the equilibrium position of the meta-atoms. The assembly of these meta-atoms for a given $a$ and $b$ gives rise to various band gaps within the frequency spectrum. By changing the values of $a$ and $b$, we can change the relative position of pass bands and band gaps. Each one of these designs (with a given $a$ and $b$) can work as a designated channel for specific frequency phonons, while rejecting the phonons within the band gap frequency ranges. We connect multiple channels composed of different configurations of $a$ and $b$, allowing only certain frequencies to pass in each channel. As a demonstration of principle, we connect three channels in a T-shaped configuration to function as a demultiplexer for infrasound phonons (Fig. \ref{fig:Schematic} d). The left and right junctions of the T-shaped waveguide have out-of-sync pass and stop bands, while the vertical junction has a pass band encompassing both pass band frequencies (Fig. \ref{fig:Schematic} b). For example, when exciting the vertical junction at a frequency  $f_1$ within the band gap of the right junction, the wave propagates only through the left channel (CH.1). Similarly, when exciting the vertical junction at a frequency $f_2$ within the band gap of the left junction, the wave propagates only through the right channel (CH.3) (Fig. \ref{fig:Schematic} c). 

\begin{figure}[b]
	\begin{center}
		\includegraphics{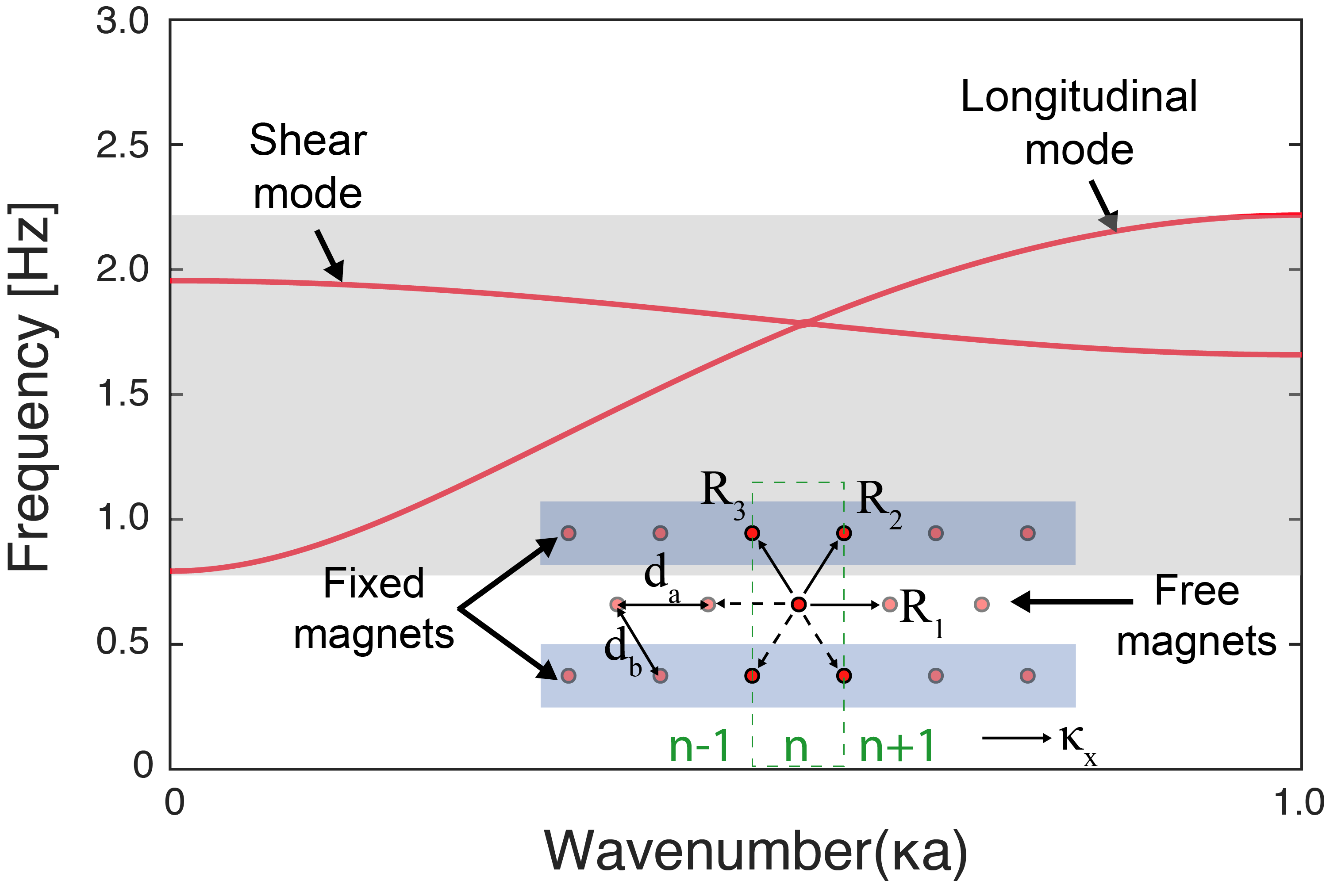}
	\end{center}
	\vspace*{-5mm}
	\caption{\textbf{Dispersion curves.} Numerical dispersion curves for both the shear and longitudinal modes for a lattice with $a=15.4~mm$, $b=24.5~mm$ and $m = 0.118~g$. The band gaps are shaded in gray. The inset shows a unit cell $n$ along with its lattice vectors.}
	\label{fig:Unit_cell}
\end{figure}

To analyze our magnetic lattices, we first consider the basic building block of our structure, a single meta-atom and its coupling to the boundary. We use Bloch's theorem to generate the dispersion curves of our magnetic lattice.\cite{bloch1929quantenmechanik} The dispersion analysis of our unit cell takes into account the nearest neighbour interaction between the meta-atoms (Fig. \ref{fig:Unit_cell} inset). Each meta-atom (i.e., the disk with the embedded concentric magnet) is coupled to four fixed boundary magnets and one other freely moving meta-atom on each side. \textcolor{black}{Each meta-atom has two degrees of freedom (i.e., movement in $x$ and $y$ direction).} The dispersion equation for our system can be written as: \begin{equation}
    [-\omega^2\textbf{M}+\textbf{K}(\boldsymbol\kappa)] \boldsymbol{\phi} = 0,
    \label{eqn:eigen}
\end{equation}
where $\omega$ is the frequency, $\kappa$ is the wavenumber, $\boldsymbol\phi = [u~v]^{T}$ is the Bloch displacement vector in $x$ and $y$ direction, $\boldsymbol {M} = \begin{bmatrix}
m & 0\\
0 & m
\end{bmatrix}$ is the mass matrix and $\boldsymbol{K}$ is the stiffness matrix\cite{jiao2019distinctive}: 
\begin{multline}
\mathbf{K}(\boldsymbol\kappa)=2\displaystyle \sum_{i=1}^{3}\left \{  {f_{,d}(d_i)\boldsymbol{e_i}\otimes \boldsymbol{e_i}[\delta_{i1}cos(\boldsymbol\kappa\cdot \boldsymbol{R_i})-1]}\right \} \\+2\displaystyle \sum_{i=1}^{3}\left \{ {\frac{f(d_i)}{d_i}(\boldsymbol{I}-\boldsymbol{e_i}\otimes \boldsymbol{e_i})[\delta_{i1}cos(\boldsymbol\kappa\cdot \boldsymbol{R_i})-1]} \right \},
\label{eqn:stiff}
\end{multline} where $d_1 = d_a$ is the distance between the meta-atoms, $d_{2,3} = d_b$ is the distance between the meta-atoms and the boundary, $\delta$ is the Kronecker delta function and $\otimes$ is the dyadic product. The stiffness matrix in equation \ref{eqn:stiff} takes into account the repulsive forces between the magnetic particles following an inverse power law in the form $f(d) = A d^{\gamma}$ and $f_{,d}(d)$ as its first derivative. For a dipole-dipole interaction\cite{mehrem2017nonlinear} $A = 3\mu B^2/4\pi$, where $\mu$ is the permeability of air and $B$ is the magnetic moment.  $\boldsymbol{R_1} = d_a  \boldsymbol {e_1}$, $\boldsymbol{R_2} = d_b \boldsymbol {e_2}$ and $\boldsymbol{R_3} = d_b \boldsymbol {e_3}$ are the lattice vectors and $\boldsymbol {e_1} = [1~~0]^{T}$, $\boldsymbol {e_2} = [d_a/2d_b~~b/2d_b]^{T}$, and $\boldsymbol {e_3} = [-d_a/2d_b~~b/2d_b]^{T}$ are the unit vectors. 

\begin{figure*}
	\begin{center}
		\includegraphics[width = \textwidth]{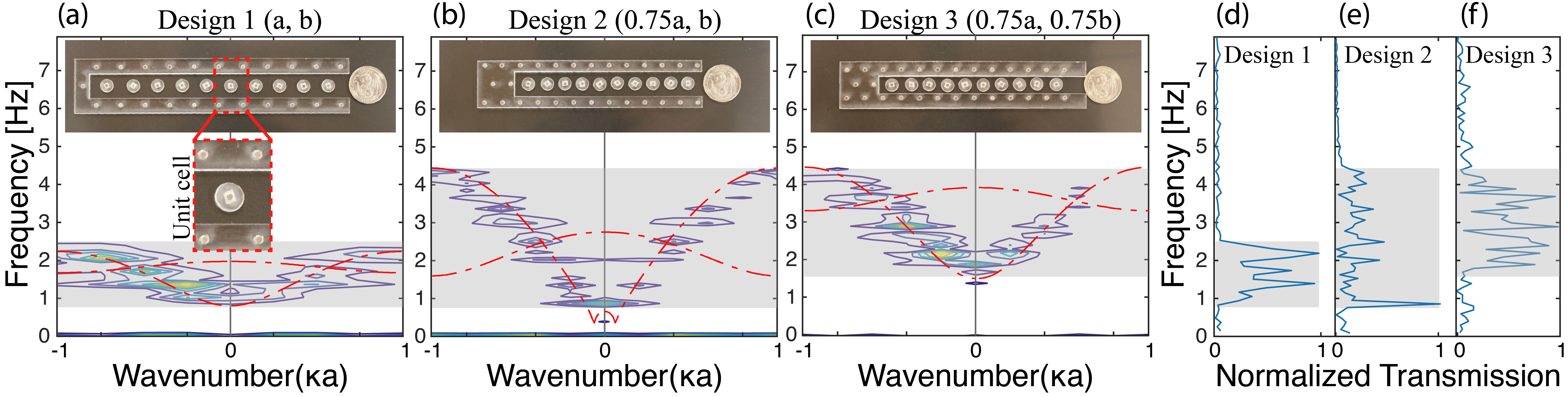}
	\end{center}
	\vspace*{-5mm}
	\caption{\textbf{Characterization of individual channels.} The experimentally measured dispersion curves plotted as contours with the numerical dispersion curves superimposed as dashed red lines for a) design 1 with $a = 15.4~mm$, $b = 24.5~mm$  b) design 2, with $a = 11.5~mm$, $b = 24.5~mm$, and c) design 3, with $a = 11.5~mm$, $b = 18.4~mm$.  The insets show the experimental samples used. The measured normalized transmission at the center of lattice for d) design 1, e) design 2, and f) design 3. The transmission bands are shaded in gray.}
	\label{fig:Disp_exp}
\end{figure*}

The dispersion curves for an example magnetic lattice with $a = 15.4~mm$, $b = 24.5~mm$, $\gamma = -4$, $m = 0.118~g$ and $A = 1.0935 \times 10^{-12}$  shows a pass band sandwiched between two band gaps (Fig. \ref{fig:Unit_cell}). The dispersion analysis considers both longitudinal and shear modes. \textcolor{black}{The two branches within the dispersion curves correlate with the two degrees of freedom of the meta-atoms. Both branches correlate with pure modes, due to the diagonal nature of the eigenvalue problem}. The pass band in figure \ref{fig:Unit_cell} shows both modes overlapping. The lower band gap is a result of the boundary coupling, while the top band gap is the cut off frequency for the lattice.  The width of the lower band gap can be reduced to zero in the limit of a vanishing boundary coupling (e.g., large $b$). The width of the pass band and its position within the frequency spectrum is more sensitive to the inter-coupling between the meta-atoms than their coupling with the boundary (i.e., the lattice constant $a$). \textcolor{black}{A larger influence of the lattice constant $a$ stems from the fact that a change in $a$ changes both $d_a$ and $d_b$ in Fig. 2 inset), while a change in the value of $b$ only changes $d_b$.} In addition to the geometric parameters of the lattice, the position of the pass band can be dynamically tuned by nondestructive external factors such as an external magnetic field.\cite{matar2013tunable,bilal2017bistable,bilal2017reprogrammable,palermo2019tuning,wang2018observation,wang2020tunable}

The purpose of the demultiplexer is to allow for the propagation of waves through the input channel, while selectively attenuating waves based on their frequencies through the output channels. To achieve such functionality, we design an input channel with a pass band that encompasses all operational frequencies, while each output channel has a pass band in only one of the operational frequency ranges. Our demultiplexer channels are designed by utilizing identical meta-atoms with varying boundary coupling $b$ and lattice spacing $a$. The modification of $a$ and $b$ allows for the control of the band gap position within the frequency spectrum. Three designs are created; design 1 with dimensions $a$ and $b$, design 2 with $0.75a$ and $b$ and design 3 with $0.75a$ and $0.75b$. The dispersion curves for each design are plotted with dashed red lines in (Fig. \ref{fig:Disp_exp} a, b, and c) based on equation \ref{eqn:eigen}. 

To experimentally verify the analytically computed dispersion curves, we fabricate three separate U-shaped boundaries with the parameters of designs 1-3 out of acrylic glass using a laser cutter (Full-spectrum 24 pro-series). The meta-atoms (i.e., disks) are identical across all three designs with a radius $r_{disk} = 4$ mm. The meta-atoms are placed within the boundary and the magnets oriented with the north pole facing upwards. The fixed magnets inside the boundary are oriented in the same way. To minimize friction, we attach a glass slide at the bottom of each disk and float them on an air bearing (New way S1030002). Each design is excited separately with a chirp signal between 0.1 Hz and 10 Hz  at its open end with a mechanical shaker (Br\"{u}el and Kj\ae r 4180) and a function generator (Keysight Technologies 33512B). The motion of the disks is captured using a computer vision camera (Blackfly S USB3) and the resulting images are analyzed using the digital image correlation software (DICe). 

The displacement profiles of all the meta-atoms in each design (characterized separately) are post-processed using 2D fast Fourier transform (2D-FFT). The resulting FFT corresponds to the longitudinal dispersion curve of each design, as we only consider the x-displacement in our analysis (Fig. \ref{fig:Disp_exp}). \textcolor{black}{For all three cases, the measured displacement signals through DICe show displacements in the y-direction on the same order as noise}. The experimental dispersion curves for the three designs match well with the numerically calculated ones. In addition, we plot the measured transmission at a central meta-atom in each design. The pass band region within the transmission spectrum matches well with the dispersion curves for all 3 designs. Design 1 has a pass band between $0.8$ Hz and $2.5$ Hz, while design 3 has a pass band between $1.6$ Hz and $4.5$ Hz. Both designs have operational frequency ranges that do not overlap (e.g., below $1.6$ Hz and above $2.5$ Hz). Design 2 has a pass band between $0.8$ Hz and $4.5$ Hz, which encompasses the pass bands of both design 1 and 3. This pass band range gives channel 2 the characteristic to act as an input channel for our demultiplexer.

\begin{figure}[b]
	\begin{center}
		\includegraphics{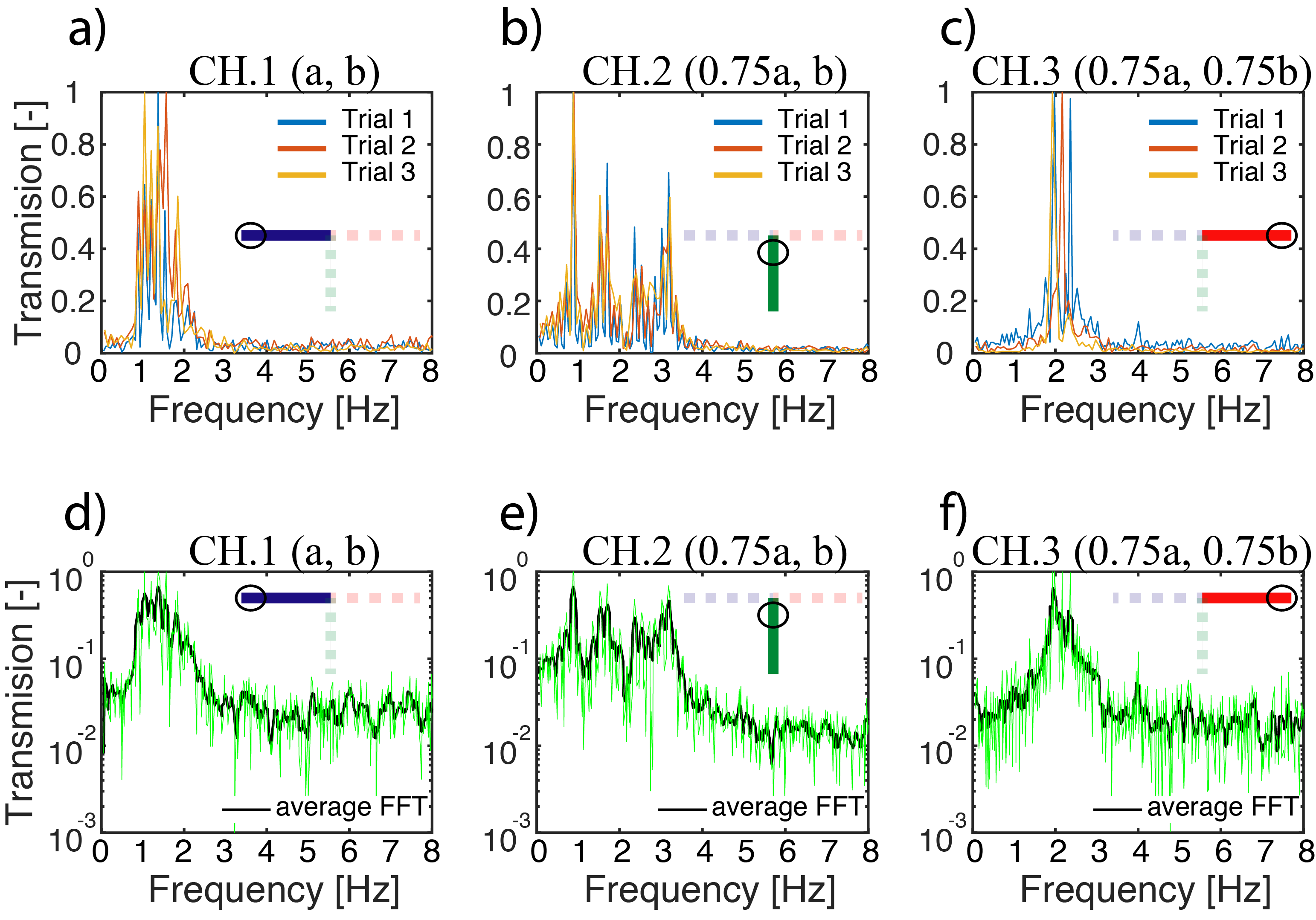}
	\end{center}
		\vspace*{-3mm}
	\caption{\textbf{Chirp signal transmission through the demultiplexer.} Fast Fourier transform of the displacement of the central meta-atom and their average over 3 trials for a,d) channel 1, b,e) channel 2, and c,f) channel 3. The input channel is excited with a chirp signal between $0.1-10$ Hz. The inset shows the corresponding channel in the T-shaped demultiplexer.}
	\label{fig:FFT_demx}
\end{figure}

\begin{figure*}
	\begin{center}
		\includegraphics[width = \textwidth]{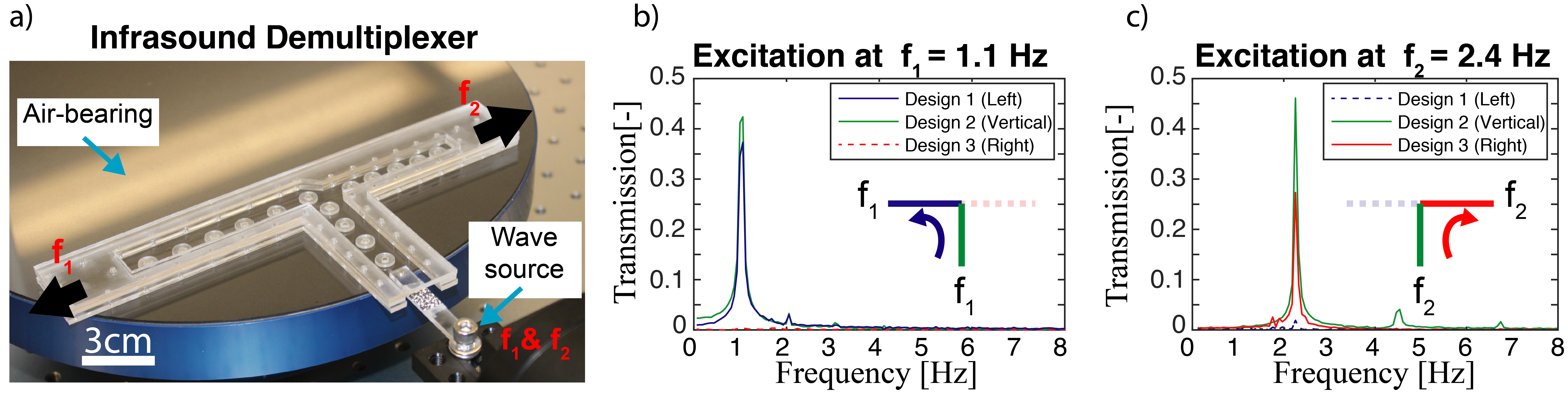}
	\end{center}
		\vspace*{-5mm}
	\caption{\textbf{Infrasound demultiplexer.} a) The experimental setup for the T-shaped demultiplexer positioned on an air bearing. The demultiplexer is filled with identical meta-atoms confined between different magnetic boundaries. A mechanical shaker drives the input channel through an arm with an embedded magnet. The signal transmission at b) $f_1=1.1$ Hz and c) $f_2=2.4$ Hz.}
	\label{fig:device}
\end{figure*}

To test our design principle we connect the three distinct channels made out of designs 1, 2 and 3. We note that the meta-atoms behave as a self-assembled fluidic material. Therefore, once the three designs are merged within a T-junction, the meta-atoms' equilibrium positions are altered. To ensure that the position of the meta-atom within each unit cell is as close as possible to the equilibrium position as in the separate experiments, we add an extra meta-atom and a boundary magnet at the junction point. To verify the operational frequencies of the T-shaped assembly, we excite the input channel (design 2) with a chirp signal from $0.1$ Hz to $10$ Hz. A fast Fourier transform is performed on the displacements of the central meta-atoms in each channel and repeated for a total of three trials (Fig. \ref{fig:FFT_demx} a, b, and c). The FFT of the three trials along with its average, for each channel, is plotted in log-scale (Fig. \ref{fig:FFT_demx} d, e, and f). The transmission of channel 1 is between $0.9$ Hz and $2.3$ Hz, for channel 2 is between $0.8$ Hz and $3.4$ Hz, and for channel 3 is between $1.7$ Hz and $2.7$ Hz. The operational frequency of each channel is altered slightly. This may be due to the imbalance in the force potentials between the different channels which causes a shift in the equilibrium position of each meta-atom.  

To characterize the performance of the demultiplexer at a single operational frequency, we excite the input channel with two sine waves, separately, and measure the transmission at each output terminal (Fig. \ref{fig:device} a). The device is first excited at a frequency $f_1=1.1$ Hz, corresponding to an operational frequency for channel 1 ( left). The FFT of the displacement of the central meta-atoms in each channel shows a clear transmission of the wave from CH.2 to CH.1, while the wave is attenuated at CH.3 (Fig. \ref{fig:device} b). When excited at a frequency $f_2=2.4$ Hz, the wave is transmitted through CH.3, while attenuated in CH.1 (Fig. \ref{fig:device} c). The ratio of transmission $(T)$ at $f_1$ is 0.88, calculated as $T_{CH1}/T_{CH2}$, and the ratio of transmission at $f_2$ is 0.59, calculated as $T_{CH3}/T_{CH2}$. The ratio of transmission for the attenuated signals at CH.1 and CH.3 are 7\% and 1\%, respectively. \textcolor{black}{We attribute the relatively lower transmission ratio through channel 3 (in comparison to channel 2) to the coupling asymmetry at the junction point between the three channel. We note the existence of the second and the third harmonic of the excitation frequency in figure \ref{fig:device} b-c, due to the inherent nonlinearity in magnetic lattices. Such nonlinear response can be harnessed by increasing the amplitude of the excitation and introducing defects within the lattice. \cite{jiao2020nonlinear,moleron2019nonlinear,mehrem2017nonlinear,chong2020nonlinear,boechler2011bifurcation,deng2018metamaterials}}

In conclusion, we utilize tunable magnetic lattices to control the propagation of infrasound waves. We tune the wave propagation characteristics of these lattices by changing their boundary couplings while using the same meta-atoms. Different designs are combined to create a T-shaped demultiplexer consisting of one input channel and two output channels. The device is characterized experimentally, validating both dispersion curves and transmission frequency spectrum. This platform can be utilized to manipulate ultra-low frequency waves, within a relatively small space. The inherent nonlinear potentials between the meta-atoms can be harnessed to demonstrate phenomena with no linear parallel such as amplitude dependent response, bifurcation, chaos and solitons.\cite{mehrem2017nonlinear,porter2015granular,gendelman2018introduction, kim2019wave,ramakrishnan2020transition,amendola2018tuning,grinberg2020nonlinear} In addition, the self-assembly of the meta-atoms can be key in creating re-programmable materials with exceptional properties.\cite{culha2020statistical}


\bibliography{Multiplexer}

\end{document}